\documentstyle[12pt]{article}
\def\one{1\hskip-.37em 1}     
\begin{document}
\begin{titlepage}
\begin{centering}

\vspace{2cm}

{\Large\bf Projection Operator Approach to}\\
\vspace{0.5cm}
{\Large\bf Constrained Systems}\\

\vspace{4cm}

Jan Govaerts\footnote{E-mail address: {\tt govaerts@fynu.ucl.ac.be}}

\vspace{0.5cm}

{\em Institut de Physique Nucl\'eaire}\\
{\em Universit\'e catholique de Louvain}\\
{\em B-1348 Louvain-la-Neuve, Belgium}\\

\vspace{2cm}

\begin{abstract}

\noindent Recently, within the context of the phase space coherent state
path integral quantisation of constrained systems, John Klauder introduced a
reproducing kernel for gauge invariant physical states, which involves
a projection operator onto the reduced Hilbert space of physical states,
avoids any gauge fixing conditions, and leads to a specific measure
for the integration over Lagrange multipliers. Here, it is pointed out
that this approach is also devoid of any Gribov problems and always
provides for an effectively admissible integration over all gauge
orbits of gauge invariant systems. This important aspect of
Klauder's proposal is explicitly confirmed by two simple examples.

\end{abstract}

\end{centering} 

\vspace{6cm}

\noindent UCL-IPN-96-P03\\
hep-th/9606007\\
May 1996

\end{titlepage}

\setcounter{footnote}{0}

\section{Introduction}
\label{Sect1}

In a recent paper\cite{John1}, John Klauder considered the quantisation
of constrained systems within the context of phase space coherent
states\cite{Coherent}, reaching an important conclusion with regards
to the path integral measure for the Lagrange multipliers which are usually
introduced in order to enforce constraints. Klauder's approach does not
require gauge fixing conditions for first class constraints,
nor Dirac brackets to reduce for second class constraints, thereby avoiding
the otherwise necessary consideration of potential Gribov 
problems\cite{Gribov,Singer} or loss of manifest covariance under specific
symmetries of the system, as well as the introduction of $\delta$-functionals
and functional determinants
into path integral representations. These latter issues are characteristic
of the conventional approaches\cite{Gov4,Hen1} to the quantisation 
of constrained systems, namely Faddeev's reduced phase space 
approach\cite{Faddeev}, Dirac's quantisation\cite{Dirac} 
or the powerful BFV-BRST methods\cite{FV}.
Nevertheless, by construction, Klauder's approach must lead to gauge invariant
observables to which each of the gauge equivalence classes of
the possible configurations of the system can only
contribute once and only once. This is to be constrasted with the
situation in the conventional approaches for which such a result is
achieved only for ``admissible" gauge fixing conditions---which, in the
generic case, cannot be found\cite{Singer}---, while
observables, even though gauge invariant, do depend on the gauge equivalence
class of gauge fixing conditions which is selected through a specific choice 
of gauge fixing conditions\cite{Gov1,Gov2,Gov3,Gov4}. Consequently, it is only 
for the gauge equivalence class of admissible gauge fixing conditions 
that the correct gauge invariant result is obtained in the
conventional approaches\cite{Gov4}.

It is obviously important to provide explicit examples confirming Klauder's
proposal for the reproducing kernel or propagator 
of phy\-si\-cal gauge invariant states.
This may be done by comparing the expressions to which the proposal leads
to well established results in the case of some
constrained systems. 
Klauder's analysis\cite{John1} emphasizes the path integral representation of
quantum amplitudes using phase space coherent states.
In the present letter,
the conclusions reached in Ref.\cite{John1} are abstracted from
the specific context of phase space coherent states, 
and are considered from the operator point of view.
Specifically, the fact that Klauder's approach avoids the necessity of
gauge fixing but nevertheless leads to the gauge invariant
results associated to what would be an admissible choice of gauge fixing 
in the conventional approaches---whether such a choice is possible or not---, 
is checked explicitly by way of two simple
examples, for which fully satisfactory results are obtained. 
Such a conclusion, which must hold in general, 
is only implicit in Ref.\cite{John1}. 

The outline of the letter is as follows. In the next section, Klauder's point
of view is briefly described in the operator context.
Sects.\ref{Sect3} and \ref{Sect4} then apply the general discussion
to two examples in Minkowski spacetime, namely the free relativistic 
scalar particle and pure Yang-Mills theory in 0+1 dimensions.
Finally, some additional comments are presented in Sect.\ref{Sect5}.

\section{Physical Projector and Physical Propagator}
\label{Sect2}

Klauder's construction involves a projection operator ${\cal E}$ onto
the subspace of states annihilated by the constraints, namely the 
reduced Hilbert space of physical states. This operator
may be abstracted from the coherent state approach used by Klauder
in the following way. Although the discussion can be extended to more
general situations\cite{John1}, for the sake of simplicity let us
consider a constrained system with Grassmann even degrees of
freedom and first class constraints only whose algebra 
is closed\cite{Gov4,Hen1}. Phase space degrees of freedom
$(q^n,p_n)$ take values over the entire real line and possess
the canonical Poisson bracket structure. The closed algebra
of the first class constraints $\phi_\alpha(q,p)$, together with
the first class Hamiltonian $H_0(q,p)$, is given by,
\begin{equation}
\left\{\phi_\alpha(q,p),\phi_\beta(q,p)\right\}=
{C_{\alpha\beta}}^{\gamma}\,\phi_\gamma(q,p)\ \ \ ,\ \ \
\left\{H_0(q,p),\phi_\alpha(q,p)\right\}={C_\alpha}^\beta\,
\phi_\beta(q,p)\ \ \ .
\label{eq:gaugealgebra}
\end{equation}
Here, ${C_{\alpha\beta}}^\gamma$ and ${C_\alpha}^\beta$ are specific
structure coefficients which determine the closed algebra of connected
local Hamiltonian gauge transformations of the system.

Consequently, time evolution of the system follows from the
first order action,
\begin{equation}
S=\int dt\,\left[\,\dot{q}^np_n\,-\,H_T(q,p)\,\right]\ \ \ ,
\end{equation}
the total Hamiltonian being given by,
\begin{equation}
H_T(q,p;\lambda)=H_0(q,p)\,+\,\lambda^\alpha\,\phi_\alpha(q,p)\ \ \ ,
\end{equation}
where the quantities $\lambda^\alpha(t)$ are arbitrary time dependent
Lagrange multipliers for the first class constraints. These Lagrange 
multipliers parametrise the local Hamiltonian gauge freedom of the system 
associated to the constraints. In particular, local Hamiltonian 
gauge transformations are given by,
\begin{equation}
\begin{array}{r c l}
\delta_\epsilon\,q^n&=&\left\{q^n\,,\,\phi_\epsilon(q,p)\,\right\}\ \ \ ,\\ \\
\delta_\epsilon\,p_n&=&\left\{p_n\,,\,\phi_\epsilon(q,p)\,\right\}\ \ \ ,\\ \\
\delta_\epsilon\,\lambda^\alpha&=&\dot{\epsilon}^\alpha\,+\,
\lambda^\gamma\epsilon^\beta\,{C_{\beta\gamma}}^\alpha\,-\,
\epsilon^\beta\,{C_\beta}^\alpha\ \ \ ,
\end{array}
\end{equation}
where the gauge generator is defined in terms of infinitesimal functions
$\epsilon^\alpha(t)$ by the com\-bi\-na\-tion
$\phi_\epsilon(q,p)=\epsilon^\alpha\,\phi_\alpha(q,p)$.
These transformations provide the basis for an analysis of the space
of gauge orbits of the system in its Hamiltonian formulation, and 
for a discussion of the possibility of admissible gauge fixing conditions,
or otherwise, of Gribov problems either of the first or second type,
or both\cite{Gov4}. Such issues must be addressed on a case by case basis.

Let us now consider the quantised system. Namely, let us assume that a choice
of quantum operator ordering and of inner product on the space of states
is possible such that the quantum algebra of constraints among themselves
and with the Hamiltonian retains the same form 
(\ref{eq:gaugealgebra}) as at the classical level,
and such that quantum observables obey the appropriate
self-adjoint properties. With the quantum system defined from its
classical counterpart in this manner, physical or gauge invariant states---at
least invariant under those gauge transformations continuously 
connected to the identity transformation---are defined by the condition,
\begin{equation}
\hat{\phi}_\alpha\,|{\rm physical}>=0\ \ \ .
\end{equation}
Time evolution of the system is induced by the total quantum
Hamiltonian $\hat{H}_T$ via the time-ordered propagator\footnote{Units such that
$\hbar=1$ are assumed throughout.},
\begin{equation}
S(t_2,t_1)=Te^{-i\int_{t_1}^{t_2}dt\hat{H}_T}\ \ \ ,
\end{equation}
which thus involves the arbitrary time dependent functions 
$\lambda^\alpha(t)$. Obviously, the action of the total {\sl Hamiltonian\/}
$\hat{H}_T$ on any physical state leads to another physical state which is independent
of the choice of Lagrange multipliers $\lambda^\alpha(t)$, this being 
not necessarily a property shared
by the propagator itself. Consequently, the evolution operator 
$S(t_2,t_1)$ does also propagate gauge {\sl variant\/} or unphysical states
in a gauge dependent manner.

In order to construct a propagator for physical states only, Klauder considers
the projection operator onto the subspace of states annihilated
by the first class quantum constraints.
Denoting this operator by ${\cal E}$, with the properties,
\begin{equation}
{\cal E}^2={\cal E}\ \ \ ,\ \ \ {\cal E}^\dagger={\cal E}\ \ \ ,
\label{eq:Proj1}
\end{equation}
the physical projector is given by\cite{John1}\footnote{When the spectrum
of the constraints $\hat{\phi}_\alpha$ is continuous, a proper definition
of the reduced physical Hilbert space requires some form of the
$\delta$-limiting procedure discussed in Ref.\cite{John1}.},
\begin{equation}
{\cal E}=\int\,dU(\theta^\alpha)\,e^{-i\theta^\alpha\hat{\phi}_\alpha}\ \ \ ,
\end{equation}
where $dU(\theta^\alpha)$ is a suitable integration measure over the space
of transformations ge\-ne\-ra\-ted by the first class constraints, such
that ${\cal E}$ does possess the properties in (\ref{eq:Proj1}).
In particular, note how the condition ${\cal E}^2={\cal E}$ 
determines the normalisation of the integration measure $dU(\theta^\alpha)$. 
For example, if these constraints generate a compact Lie group, 
$dU$ is the associated
normalised Haar measure over that group\footnote{To be precise, first class
constraints generate only the connected component of the Lie group,
whereas the full gauge group of the system may be different from its universal
covering group. Such a situation may properly be implemented by appropriatedly
modifying the integration domain over the group parameters $\theta^\alpha$
in the definition of the projector ${\cal E}$.}.

Given the physical projector ${\cal E}$, the {\sl physical propagator\/} 
for gauge invariant states is then constructed 
to be\cite{John1}\footnote{Note that this construction is reminiscent
of Feynman's tree theorem\cite{Feynman}. Similar or somewhat different types
of projections may also be found in Refs.\cite{Hen1,Hajicek}.},
\begin{equation}
S_{\rm phys}(t_2,t_1)=e^{-i\hat{H}_0(t_2-t_1)}\,{\cal E}\ \ \ .
\label{eq:Sphys}
\end{equation}
Since the first class constraints form a closed algebra among themselves
and with the canonical Hamiltonian $\hat{H}_0$, note that one may also write,
\begin{equation}
S_{\rm phys}(t_2,t_1)={\cal E}\,e^{-i\hat{H}_0(t_2-t_1)}\,{\cal E}=
{\cal E}\,e^{-i{\cal E}\hat{H}_0{\cal E}(t_2-t_1)}\,{\cal E}\ \ \ .
\end{equation}
In particular, the latter of these two expressions is the one relevant
more generally for systems which include second class constraints
as well\cite{John1}. In this form, it should be clear that the physical
propagator does indeed propagate as intermediate states physical states
only, and as external states their gauge invariant components only.
Moreover, this propagator obeys the convolution property required
of an evolution operator,
\begin{equation}
S_{\rm phys}(t_3,t_2)\,S_{\rm phys}(t_2,t_1)=S_{\rm phys}(t_3,t_1)\ \ \ .
\end{equation}

Once the choice of physical evolution operator is specified, it is possible
of course to compute its matrix elements for different choices
of quantum states. The latter may include for example configuration
space eigenstates, momentum space eigenstates, or phase space coherent
states. Whatever the choice, it is then also possible to develop
a path integral representation of such matrix elements in the usual
manner, by inserting resolutions of the identity operator $\one$
in terms of the chosen set of states in a step-wise discretised version
of the evolution operator. Since the quantised system is
assumed to have been completely defined at the operator level,
including the projection
operator ${\cal E}$, one obviously obtains a path integral representation
in which the measure of all phase space degrees of freedom
and Lagrange multiplier variables is uniquely determined and well defined.
In particular, the property ${\cal E}^2={\cal E}$
in (\ref{eq:Proj1}) of the physical
projector ${\cal E}$ uniquely
determines the integration measure over the Lagrange multipliers in
a path integral representation of matrix elements of the physical
evolution operator\cite{John1}.

This is achieved in spite of the absence of any choice
of gauge fixing, thereby avoiding any potential Gribov problems
in the evaluation of quantities which are gauge invariant observables
by construction. Indeed in the conventional approaches, even though
gauge fixing can be effected in a manner which necessarily
ensures the gauge invariance 
of expressions, nevertheless it leads to results which do depend on 
the gauge equivalence class to which the chosen gauge fixing conditions belong.
As pointed out previously, it is only for admissible gauge fixing conditions
that physically consistent results are obtained for gauge invariant observables.

In contradistinction, the choice of physical evolution operator 
in (\ref{eq:Sphys})
avoids any such difficulties at once. No choice of gauge fixing condition
is to be effected, hence no issue of a possible Gribov problem can arise.
Nevertheless, gauge invariant results are obtained, owing to 
the physical projector ${\cal E}$,
by properly integrating over the space of gauge transformations.
Moreover, not only does one obtain gauge invariant results, but
in addition these results must necessarily be such as to
include properly once and only once 
the contribution of each of the gauge inequivalent configurations of the system.
There is no need to go into the development of a BRST invariant
approach in order to maintain a formulation
of the system which is both at the same time manifestly gauge invariant
and covariant under other specific symmetries.

Ref.\cite{John1}, emphasizing the path integral point of view within the
phase space coherent state approach, illustrated
through a series of examples how the projector property of the operator 
${\cal E}$ does indeed determine the path integral measure 
over the Lagrange multipliers. In the present letter, and within the abstract
operator approach, it is the absence of Gribov problems and the
admissibility of the effective integration over the space of gauge orbits of
such gauge invariant systems which are pointed out, 
and illustrated explicitly by way of two simple examples. Indeed, these
important facts must again result from the properties of the physical
projector ${\cal E}$.

\section{The Relativistic Scalar Particle}
\label{Sect3}

Consider the free relativistic scalar particle of mass $m\ge 0$
propagating in a
Minkowski spacetime of $D$ dimensions.
The manifestly reparametrisation invariant Hamiltonian formulation
of this system is well known\cite{Gov5}. Using the notations and 
spacetime metric conventions of Ref.\cite{Gov4}, with in addition 
a choice of units such that $c=1$, the canonically conjugate degrees
of freedom of the system are the spacetime coordinates $x^\mu(\tau)$
and energy-momentum $P^\mu(\tau)$ $(\mu=0,1,\cdots,D-1)$
of the particle, the canonical Hamiltonian
$H_0$ vanishes identically as befits a reparametrisation invariant
dynamics, and the first class constraint related to the connected
gauge invariance of the system under orientation preserving
reparametrisations of the world-line coordinate $\tau$ is,
\begin{equation}
\phi=\frac{1}{2}\,\left[P^2+m^2\right]\ \ \ .
\end{equation}
Consequently, the total Hamiltonian of the system is simply,
\begin{equation}
H_T=\lambda\,\phi=\frac{1}{2}\lambda\left[P^2+m^2\right]\ \ \ ,
\end{equation}
where $\lambda(\tau)$ is the Lagrange multiplier associated to the
connected Hamiltonian gauge freedom generated by $\phi$.

It may be shown\cite{Gov4} that the space of gauge inequivalent configurations
of the system is characterised by the world-line metric
Teichm\"uller parameter $\gamma$ defined by,
\begin{equation}
\gamma=\int_{\tau_1}^{\tau_2}\,d\tau\,\lambda(\tau)\ \ \ ,
\label{eq:gamma}
\end{equation}
where the interval $[\tau_1,\tau_2]$ is related to a choice of
boundary conditions. In particular, the parameter $\gamma$
is invariant under the orientation preserving
reparametrisations of the world-line,
{\sl i.e.} the connected gauge transformations of the system,
generated by the first class constraint $\phi$.
Under orientation reversing reparametrisations however, the
Teichm\"uller parameter changes sign. Therefore, when describing
the oriented scalar particle invariant under both classes of
transformations, corresponding to a particle distinct from
its antiparticle, the Teichm\"uller parameter must be restricted
to a fundamental domain of the modular group\cite{Gov4}, say the interval
$\gamma\in[0,+\infty[$\ .

Quantisation of this system is straightforward enough. 
One has the fundamental operator degrees of freedom $\hat{x}^\mu$ and
$\hat{P}_\mu$ $(\mu=0,1,\cdots,D-1)$ with the canonical
commutation relations,
\begin{equation}
\left[\hat{x}^\mu\,,\,\hat{P}_\nu\right]=i\delta^\mu_\nu\ \ \ .
\end{equation}
The first class quantum constraint is simply,
\begin{equation}
\hat{\phi}=\frac{1}{2}\,\left[\hat{P}^2+m^2\,\right]\ \ \ \ ,
\end{equation}
while the generator of time evolution is the total quantum Hamitonian,
\begin{equation}
\hat{H}_T=\lambda\,\hat{\phi}=
\frac{1}{2}\,\lambda\,\left[\hat{P}^2+m^2\right]\ \ \ .
\label{eq:HT}
\end{equation}

Since the first class Hamiltonian $\hat{H}_0$ vanishes identically
for this system, the proposal of Ref.\cite{John1} corresponds
to the statement that the physical time evolution operator of the system
is simply the projection operator ${\cal E}$, which in the present case
is defined by,
\begin{equation}
S_{\rm phys}(\tau_f,\tau_i)={\cal E}=\,\int_{-\infty}^{+\infty}\,d\gamma\,
e^{-\frac{1}{2}i\gamma(\hat{P}^2+m^2)}\,
\frac{\sin(\delta\gamma)}{\pi\gamma}\ \ \ ,
\ \ \ 0<\delta <<1\ \ \ ,
\label{eq:Proj2}
\end{equation}
with a suitable $\delta\rightarrow 0$ limit reserved to a later 
stage\cite{John1}. Note how the integration parameter $\gamma$ 
is indeed to be identified with the Teichm\"uller parameter of the system 
defined in (\ref{eq:gamma}), on basis of the total Hamiltonian 
in (\ref{eq:HT}). {\sl A priori\/}, the integration
measure over the parameter $\gamma$ could be any function of $\gamma$,
since $\gamma$ is invariant under local world-line reparametrisations.
However, the requirements in (\ref{eq:Proj1}) necessary for
a projection operator imply in fact that the integration measure over
$\gamma$ be precisely of the form as specified in (\ref{eq:Proj2})
for some $\delta >0$. In other words, the requirement that ${\cal E}$ 
be a projection operator essentially onto the sector of physical---or 
locally gauge invariant---states effectively determines 
the integration measure over Teichm\"uller and modular space.

Given the desired projection operator, its matrix elements are
computable in a straightforward manner. Let us first consider
the configuration space matrix elements, namely,
\begin{equation}
P(x^\mu_i\rightarrow x^\mu_f)\equiv <x^\mu_f|{\cal E}|x^\mu_i>\ \ \ ,
\end{equation}
where the states $|x^\mu>$ define the complete orthonormalised 
basis of eigenvectors of
the position operators $\hat{x}^\mu$. A similar orthonormalised
basis of momentum eigenstates $|p_\mu>$ exists for the momentum operators
$\hat{P}_\mu$. These two bases are related through the transformation rule,
\begin{equation}
<p_\mu|x^\mu>=\left(2\pi\right)^{-D/2}\,e^{-i x\cdot p}\ \ \ ,
\label{eq:basisrule}
\end{equation}
in which the invariant inner product in the exponential is obviously
the one defined by the Minkowski metric on spacetime.
Using this rule as well as the spectral decomposition of the identity
operator $\one$ in terms of the momentum eigenstates $|p_\mu>$,
it is straightforward to obtain for the configuration space matrix elements
of the physical evolution operator,
\begin{equation}
\begin{array}{r c l}
S_F(x^\mu_i\rightarrow x^\mu_f)&\equiv&\lim_{\delta\rightarrow 0}
\frac{\pi}{2\delta}\,
P(x^\mu_i\rightarrow x^\mu_f) \\ \\
&=&\lim_{\delta\rightarrow 0}\,
\frac{\pi}{2\delta}\,\int_{(\infty)}\frac{d^Dp^\mu}{(2\pi)^D}\,
e^{i(x_f-x_i)\cdot p}\,\int_{-\infty}^{+\infty}d\gamma\,
e^{-\frac{1}{2}i\gamma(p^2+m^2)}\,\frac{\sin(\delta\gamma)}{\pi\gamma}\\ \\
&=&\frac{1}{2} \,\int_{(\infty)}\frac{d^Dp^\mu}{(2\pi)^D}\,
e^{i(x_f-x_i)\cdot p}\,\int_{-\infty}^{+\infty}d\gamma\,
e^{-\frac{1}{2}i\gamma(p^2+m^2)}\ \ \ ,
\end{array}
\label{eq:Prop1}
\end{equation}
where the limit $\delta\rightarrow 0$ is taken in the way discussed
in Ref.\cite{John1}. The choice of normalisation of the function
$S_F(x^\mu_i\rightarrow x^\mu_f)$ is such that when restricting
the modular parameter $\gamma$ to the range $[0,+\infty]$---corresponding
to the description of the oriented particle---, the
function $S_F(x^\mu_i\rightarrow x^\mu_f)$ coincides with
the Feynman propagator for the scalar particle.

Up to a constant factor, note that it is only with the integration measure 
over the pa\-ra\-me\-ter $\gamma$ which appears in (\ref{eq:Prop1}) that 
the Feynman propagator is obtained in that manner. 
Any other non constant integration measure
over $\gamma$, even though gauge invariant for local
and possibly global gauge transformations---{\sl i.e\/} for 
orientation preserving and reversing world-line reparametrisations,
respectively---, would not lead to the Feynman
propagator, and would thus introduce a Gribov problem of some
type\cite{Gov4}. In the present instance, as was pointed out above, 
it is precisely the fact that ${\cal E}$ is a projection operator 
with the properties in (\ref{eq:Proj1})
which ensures the admissible integration measure over modular space, 
devoid of any Gribov problem. In addition, the appropriate physical
propagation of gauge---{\sl i.e.\/} reparametrisation invariant---states is
indeed recovered, in spite of the fact that no gauge fixing
of the system is effected. Compared to the detailed calculation
of the physical propagator $S_F(x^\mu_i\rightarrow x^\mu_f)$ using
Hamiltonian BRST techniques\cite{Teitelboim,HenTeitel,Monaghan,Gov4},
it is clear that the projection operator approach is far more efficient
and leads immediately to the correct result, in contradistinction to the
BRST approach for which the correct result is obtained only for an admissible
choice of gauge fixing condition\cite{Gov4}.

In view of the basis for the analysis of Ref.\cite{John1}, let us also
compute the phase space coherent state\footnote{In Ref.\cite{John1}, 
the initial example of a first class constraint considers the motion 
of a particle on a hypersphere with vanishing Hamiltonian. 
Ref.\cite{John1} discusses
how the associated coherent states are related to 
the euclidian group which appears
in that context. The relativistic particle is similar in that
the constraint enforces the {\sl momentum\/} to lie on a hypersphere
of Minkowski signature. One may thus raise the question of the
characterisation of the group of transformations in momentum
space {\sl and in spacetime\/} related to the phase space
coherent state matrix elements of the evolution operator
for the relativistic particle, along the lines of the first example
in Sect.6 of Ref.\cite{John1}.} matrix
elements of the physical evolution operator ${\cal E}$. 
These coherent states are defined by,
\begin{equation}
|P_\mu,x^\mu>=e^{i\alpha(P_\mu,x^\mu)}\,
e^{-ix^\mu\hat{P}_\mu}\,e^{iP_\mu\hat{x}^\mu}\,|\eta>\ \ \ ,
\label{eq:coherent1}
\end{equation}
with an arbitrary phase factor $\alpha(P_\mu,x^\mu)$ and normalised fiducial
state $|\eta>$. It is then a simple exercise to compute
the coherent state matrix elements of the projector ${\cal E}$,
\begin{displaymath}
<P_2,x_2|{\cal E}|P_1,x_1>=e^{-i\alpha(P_2,x_2)}\,e^{i\alpha(P_1,x_1)}\times
\end{displaymath}
\begin{equation}
\times\int_{(\infty)}d^Dp^\mu\,e^{i(x_2-x_1)\cdot p}\,
\eta^*(p-P_2)\,\eta(p-P_1)\,\int_{-\infty}^{+\infty}
d\gamma\,e^{-\frac{1}{2}\gamma(p^2+m^2)}\,
\frac{\sin(\delta\gamma)}{\pi\gamma}\ \ \ ,
\end{equation}
where $\eta(P_\mu)$ is the momentum space wave function of the
fiducial state $|\eta>$, namely the quantity $\eta(P_\mu)=<P_\mu|\eta>$.

Given this expression and the resolution of the identity operator $\one$ 
in terms of the overcomplete basis of phase space coherent states, 
it is straightforward to verify that the
configuration space matrix elements of the projection operator ${\cal E}$
are again given by (\ref{eq:Prop1}), independently of the choice
of fiducial state $|\eta>$ used in the definition of coherent states.
This check uses the relation,
\begin{equation}
P(x^\mu_i\rightarrow x^\mu_f)=\int_{(\infty)}\frac{d^DP_2d^Dx_2}{(2\pi)^D}
\frac{d^DP_1d^Dx_1}{(2\pi)^D}\,<x^\mu_f|P_2,x_2>
<P_2,x_2|{\cal E}|P_1,x_1><P_1,x_1|x^\mu_i>\ \ \ ,
\end{equation}
as well as the overlap functions $<y^\mu|P,x>$ which are easily obtained from 
(\ref{eq:coherent1}) and (\ref{eq:basisrule}).

\section{Pure Yang-Mills Theory in 0+1 Dimensions}
\label{Sect4}

Let us consider a pure Yang-Mills theory in a Minkowski spacetime
of $1+1$ dimensions, based on an arbitrary simple compact Lie group
$G$ of dimension $D_G$ and of rank $\ell$. 
The Lie algebra generators $T^a$ $(a=1,2,\cdots,D_G)$
obey the commutation relations,
\begin{equation}
[T^a,T^b]=if^{abc}T^c\ \ \ ,
\end{equation}
with real, fully antisymmetric structure coefficients $f^{abc}$.
In particular, the adjoint re\-pre\-sen\-tation of dimension $D_G$ possesses
the matrix representation
$\left(T^a_{\rm Adj}\right)^{bc}=-if^{abc}$.
The gauge vector potential components are denoted $A^a_\mu$ $(\mu=0,1)$,
while the gauge coupling constant is denoted by $g$, so that the gauge
field strength is
$F^a_{\mu\nu}=\partial_\mu A^a_\nu-\partial_\nu A^a_\mu
+gf^{abc}A^b_\mu A^c_\nu$.

Given these data, let us now consider\footnote{This system is discussed
in Ref.\cite{Shabanov1} for example.} the dimensional reduction
of this pure Yang-Mills theory to $0+1$ dimensions, by retaining
only the $\partial_1$ zero modes of the fields in space, 
namely by assuming that the fields $A^a_\mu(t=x^0,x^1)$ are now 
independent of the space coordinate $x^1$.
In order to avoid any confusion, 
let us then distinguish the time and space components of 
the gauge fields as follows,
\begin{equation}
\phi^a(t)=A^a_0(t)\ \ \ ,\ \ \ A^a(t)=A^a_1(t)\ \ \ ,
\end{equation}
so that the only non vanishing component of the field strength is
now given by,
\begin{equation}
F^a_{01}=\dot{A}^a+gf^{abc}\phi^b A^c\ \ \ .
\end{equation}
 
Consequently, the dimensionally reduced system is described by the
Lagrangian,
\begin{equation}
L=\frac{1}{2}\left[\dot{A}^a+gf^{abc}\phi^b A^c\right]^2\,-\,
\frac{1}{2}m^2(A^a)^2\ \ \ ,
\label{eq:Lag}
\end{equation}
where a gauge invariant mass term for the $A^a$ degrees of 
freedom has been added. Indeed, the reduced system possesses the
following gauge invariance,
\begin{equation}
{A'}^aT^a=U\,A^aT^a\,U^{-1}\ \ \ ,\ \ \ 
{\phi'}^aT^a=U\,\phi^aT^a\,U^{-1}\,+\,\frac{i}{g}\,U\,\frac{d}{dt}U^{-1}\ \ \ ,
\end{equation}
where $U(t)=e^{-ig\theta^a(t)T^a}$ is an arbitrary time dependent
transformation in $G$. Quite obviously, the mass term does not spoil
this gauge invariance. As will become clear later on, 
the mass term serves the purpose of a regularisation of the quantised theory.

Owing to the absence of a dependence on the time
derivative of the degrees of freedom $\phi^a$ in the above Lagrangian,
the present is a constrained system possessing a gauge in\-va\-ri\-ance
under the simple compact Lie group $G$. As a matter of fact, the full
gauge invariance of the system, including connected---{\sl i.e.} local---and
non connected---{\sl i.e.} global---gauge transformations is not
the universal covering group $G_{\rm univ}$ generated by the above
algebra, but rather the simple compact Lie group $G=G_{\rm univ}/C$, 
where $C$ is the maximal torus or center of the group $G_{\rm univ}$. 
Indeed, the degrees of freedom $\phi^a$ and $A^a$ transform under the adjoint
representation of $G$ or $G_{\rm univ}$.

\vspace{10pt}

\noindent{\bf The Hamiltonian formulation}. It is straightforward 
to apply the usual analysis of constraints
starting from the Lagrangian (\ref{eq:Lag}). Details are not presented.
Let us only point out that the analysis follows the same lines\cite{Gov4} as
for Yang-Mills theory in a Minkowski spacetime of dimension $(D-1)+1$,
and that some of the degrees of freedom---namely
the sector of the coordinates $\phi^a$ and their conjugate momenta---may 
be decoupled by considering the so called\cite{Gov4} 
{\sl fundamental Hamiltonian description\/} of the system. 

In the present instance, this fundamental description is based on the
phase space degrees of freedom, that is the coordinates
$A^a(t)$ and their conjugate momenta $\pi^a(t)$, 
obeying the algebra of Poisson brackets,
\begin{equation}
\{A^a(t),\pi^b(t)\}=\delta^{ab}\ \ \ .
\end{equation}
The system possesses first class constraints only, namely the gauge
charges generating the local gauge transformations, which in fact
also enforce Gauss' law in the present case,
\begin{equation}
Q^a=gf^{abc}A^b\pi^c\ \ \ ,
\end{equation}
whose closed algebra is simply that of the Lie algebra of the gauge group $G$,
\begin{equation}
\{Q^a(t),Q^b(t)\}=gf^{abc}Q^c(t)\ \ \ .
\end{equation}
Finally, the first class Hamiltonian is simply,
\begin{equation}
H_0=\frac{1}{2}\left(\pi^a\right)^2+\frac{1}{2}m^2\left(A^a\right)^2\ \ \ ,
\end{equation}
so that the total Hamiltonian generating the time evolution of the
system is,
\begin{equation}
H_T=\frac{1}{2}\left(\pi^a\right)^2+\frac{1}{2}m^2\left(A^a\right)^2\,-\,
\phi^a Q^a\ \ \ .
\end{equation}
Here, the variables $\phi^a$ are {\sl Lagrange multipliers\/}\footnote{In fact,
the variables $\phi^a$ introduced here correspond to the opposite
of the Lagrange multipliers $\lambda^a$ introduced in Sect.\ref{Sect2}
in the general case, namely $\phi^a=-\lambda^a$.}
for the first class constraints $Q^a$, which, in fact, may be identified with
the original gauge degrees of freedom as $\phi^a=A^a_0$,
the latter thus also parametrising the local gauge freedom of the system.

Given the total Hamiltonian, the Hamiltonian equations of motion are
readily derived,
\begin{equation}
\dot{A}^a=\pi^a-gf^{abc}\phi^bA^c\ \ \ ,\ \ \ 
\dot{\pi}^a=-gf^{abc}\phi^b\pi^c-m^2A^a\ \ \ ,
\end{equation}
whose solutions thus involve the arbitrary Lagrange multipliers $\phi^a$.
Similarly, local Hamiltonian gauge transformations generated by the
first class constraints $Q^a$ read,
\begin{equation}
\begin{array}{r c l}
\delta_\epsilon A^a&=&\{A^a,Q_\epsilon\}=gf^{abc}\epsilon^b A^c\ \ \ ,\\ \\
\delta_\epsilon \pi^a&=&\{\pi^a,Q_\epsilon\}=
gf^{abc}\epsilon^b \pi^c\ \ \ ,\\ \\
\delta_\epsilon \phi^a&=&-\dot{\epsilon}^a+gf^{abc}\epsilon^b\phi^c\ \ \ ,
\end{array}
\end{equation}
with,
\begin{equation}
Q_\epsilon=\epsilon^aQ^a\ \ \ ,
\end{equation}
$\epsilon^a(t)$ being arbitrary infinitesimal functions of time.
It is a straightforward exercise to check that the first-order Hamiltonian
Lagrangian,
\begin{equation}
L_{\rm Hamilt}=\dot{A}^a\pi^a-H_T\ \ \ ,
\end{equation}
is indeed invariant under these transformations, since,
\begin{equation}
\delta_\epsilon(\dot{A}^a\pi^a)=\dot{\epsilon}^aQ^a\ \ \ ,\ \ \ 
\delta_\epsilon H_T=\dot{\epsilon}^aQ^a\ \ \ .
\end{equation}

In fact, it is possible even to determine the Hamiltonian gauge transformations
to all orders, and not only in linearised form. For this purpose, let us
define the finite gauge transformation in the group $G$,
\begin{equation}
U(t)=e^{-ig\theta^a(t)T^a}\ \ \ .
\end{equation}
The gauge transformed Hamiltonian degrees of freedom are then determined
from,
\begin{equation}
\begin{array}{r c l}
{A'}^aT^a&=&UA^aT^aU^{-1}\ \ \ ,\ \ \ {\pi'}^aT^a=U\pi^aT^aU^{-1}\ \ \ ,\\ \\
{\phi'}^aT^a&=&U\phi^aT^aU^{-1}\,+\,\frac{i}{g}U\frac{d}{dt}U^{-1}\ \ \ .
\end{array}
\end{equation}

Consequently, complete gauge fixing in this system is possible. Indeed,
consider a certain configuration for $(A^a,\pi^a,\phi^a)$ and define
the gauge transformation,
\begin{equation}
U(t,t_0)=Te^{-ig\int_{t_0}^{t}\,dt'\,\phi^a(t')T^a}\ \ \ .
\end{equation}
Then the transformed Lagrange multipliers vanish identically,
\begin{equation}
{\phi'}^a(t)=0\ \ \ ,
\end{equation}
while no additional gauge transformation exists which would leave this
last identity in\-va\-riant, given specific boundary conditions on $A^a$ 
and/or $\pi^a$.

\vspace{10pt}

\noindent{\bf Quantisation and physical states}. Let us now consider 
the quantised system. Given the expressions for $H_0$
and $Q^a$ at the classical level, the corresponding operators are
simply defined by,
\begin{equation}
\hat{H}_0=\frac{1}{2}\left[\left(\hat{\pi}^a\right)^2+
m^2\left(\hat{A}^a\right)^2\right]\ \ \ ,
\end{equation}
and
\begin{equation}
\hat{Q}^a=gf^{abc}\hat{A}^b\hat{\pi}^c\ \ \ .
\end{equation}
Due to the fundamental commutation relations,
\begin{equation}
\left[\hat{A}^a(t),\hat{\pi}^b(t)\right]=i\delta^{ab}\ \ \ ,
\end{equation}
and the complete antisymmetry of the structure coefficients $f^{abc}$,
the operators $\hat{H}_0$ and $\hat{Q}^a$ as defined above do not
suffer quantum ordering ambiguities.

In view of the analogy with the ordinary harmonic oscillator, it is useful
to introduce the Fock representation of the system, in terms of the
operators,
\begin{equation}
\alpha^a=\sqrt{\frac{m}{2}}\,\left[\hat{A}^a+\frac{i}{m}\hat{\pi}^a\right]\ \ ,
\ \ \ {\alpha^a}^\dagger=\sqrt{\frac{m}{2}}\,
\left[\hat{A}^a-\frac{i}{m}\hat{\pi}^a\right]\ \ ,
\end{equation}
or,
\begin{equation}
\hat{A}^a=\frac{\alpha^a+{\alpha^a}^\dagger}{\sqrt{2m}}\ \ \ ,\ \ \ 
\hat{\pi}^a=-i\sqrt{\frac{m}{2}}\,\left[\alpha^a-{\alpha^a}^\dagger\right]\ \ ,
\end{equation}
such that,
\begin{equation}
\left[\alpha^a,{\alpha^b}^\dagger\right]=\delta^{ab}\ \ \ .
\end{equation}

The Hamiltonian $\hat{H}_0$ then reads,
\begin{equation}
\hat{H}_0=\frac{1}{2}m
\left[\alpha^a{\alpha^a}^\dagger+ {\alpha^a}^\dagger\alpha^a\right]=
m\left[{\alpha^a}^\dagger\alpha^a+\frac{1}{2}D_G\right]\ \ \ ,
\end{equation}
while the generators of local gauge transformations become,
\begin{equation}
\hat{Q}^a=-igf^{abc}{\alpha^b}^\dagger\alpha^c\ \ \ .
\end{equation}
Obviously, one has in particular,
\begin{equation}
\left[\hat{Q}^a,\hat{Q}^b\right]=igf^{abc}\hat{Q}^c\ \ \ .
\end{equation}

Given the normalised Fock vacuum $|0>$,
\begin{equation}
\alpha^a|0>=0\ \ \ ,\ \ \ <0|0>=1\ \ \ ,
\end{equation}
the orthonormalised basis of the Fock space, spanned by
\begin{equation}
|a_1,a_2,\cdots,a_n>=
N(a_1,a_2,\cdots,a_n)\,{\alpha^{a_1}}^\dagger{\alpha^{a_2}}^\dagger\,
\cdots\,{\alpha^{a_n}}^\dagger\,|0>\ \ \ ,
\label{eq:spacestates}
\end{equation}
where $N(a_1,a_2,\cdots,a_n)$ is a normalisation factor, also diagonalises
the Hamiltonian $\hat{H}_0$ of the system, with,
\begin{equation}
\hat{H}_0|a_1,a_2,\cdots,a_n>=m(n+\frac{1}{2}D_G)|a_1,a_2,\cdots,a_n>\ \ \ .
\end{equation}
Note that this basis of orthonormalised states is in one-to-one correspondence
with all fully symmetric irreducible representations of the unitary
group $SU(D_G)$, whose Young tableaux reduce to single rows of all
possible lengths $(n=0,1,\cdots)$.

Consider now the subspace of physical states defined by the condition
of local gauge in\-va\-rian\-ce,
\begin{equation}
\hat{Q}^a|{\rm physical}>=0\ \ \ .
\end{equation}
In view of the structure of the charges $\hat{Q}^a$, it is possible to
show\cite{Shabanov1} that such states are necessarily all of the form,
\begin{equation}
|n_1,\cdots,n_\ell>=N(n_1,\cdots,n_\ell)\,
\Big[{\rm Tr}\left(\alpha^\dagger\right)^{r_1}\Big]^{n_1}\,\cdots\,
\Big[{\rm Tr}\left(\alpha^\dagger\right)^{r_\ell}\Big]^{n_\ell}|0>\ \ \ .
\label{eq:physstates}
\end{equation}
Here, $N(n_1,\cdots,n_\ell)$ are normalisation factors whose
evaluation has to be considered on a case by case basis for
every choice of gauge group $G$,
$n_1,\cdots,n_\ell$ are arbitrary positive or vanishing integers,
$r_1,\cdots,r_\ell$ are the degrees of the independent invariant
symmetric polynomials or Casimir operators in the group $G$ of
rank $\ell$, and finally, the operators $\alpha$ and $\alpha^\dagger$
are defined by,
\begin{equation}
\alpha\,=\,\alpha^a\,T^a\ \ \ ,
\ \ \ \alpha^\dagger={\alpha^a}^\dagger\,T^a\ \ \ ,
\end{equation}
the traces in (\ref{eq:physstates}) being taken in colour space only.

The orthonormalised states $|n_1,\cdots,n_\ell>$ are gauge singlets, 
as befits physical states, and span
the entire space of physical states. In addition, they also diagonalise
the Hamiltonian $\hat{H}_0$,
\begin{equation}
\hat{H}_0\,|n_1,\cdots,n_\ell>=
m\left(n_1r_1+\cdots+n_\ell r_\ell+\frac{1}{2}D_G\right)\,
|n_1,\cdots,n_\ell>\ \ \ .
\end{equation}

In the simple case of $G=SU(2)$ of rank $\ell=1$, it is straightforward
to compute the nor\-ma\-li\-sa\-tion factor $N(n_1)$, in a manner
which should be generalisable to an arbitrary group $G$. Let us introduce
the operators,
\begin{equation}
\begin{array}{r c l}
N=\sum_{a=1}^3{\alpha^a}^\dagger\alpha^a\ \ \ &,&\ \ \ N^\dagger=N\ \ \ ,\\ \\
B^\dagger=\sum_{a=1}^3{\alpha^a}^\dagger{\alpha^a}^\dagger\ \ \ &,&\ \ \ 
B=\sum_{a=1}^3\alpha^a\alpha^a\ \ \ ,
\end{array}
\end{equation}
whose algebra is simply,
\begin{equation}
\left[N,B\right]=-2B\ \ ,\ \ 
\left[N,B^\dagger\right]=2B^\dagger\ \ ,\ \ 
\left[B,B^\dagger\right]=4N+2D_G=4N+6\ \ \ .
\end{equation}
A simple calculation then leads to the following normalisation
of the basis $|n>$ of the subspace of physical states,
\begin{equation}
|n>=\left[2^n\,n!\,\prod_{j=1}^n(2j+D_G-2)\right]^{-1/2}\,
\left(B^\dagger\right)^n\,|0>\ \ \ ,
\end{equation}
which thus satisfy the relations,
\begin{equation}
<n|m>=\delta_{n,m}\ \ \ ,\ \ \ n,m=0,1,\cdots\ \ \ .
\end{equation}

In particular, this result allows one to determine the configuration space
wave function representation of physical states. These wave functions
are defined by,
\begin{equation}
\psi_n(A^a)\equiv <A^a|n>\ \ \ ,
\end{equation}
where $|A^a>$ are the configuration space orthonormalised
eigenstates of the operators $\hat{A}^a$. One then obtains,
\begin{displaymath}
\psi_n(A^a)=\left(\frac{m}{\pi}\right)^{D_G/4}\,
\left[2^n\, n!\, (2n+1)!!\right]^{-1/2}\times
\end{displaymath}
\begin{equation}
\times\left(\frac{m}{2}\right)^n\,
\left[\,\sum_{a=1}^3\left(A^a-\frac{1}{m}\frac{\partial}
{\partial A^a}\right)^2\,\right]^n\,e^{-\frac{1}{2}m(\sum_{a=1}^3A^a)^2}\ \ \ .
\label{eq:wavefunction}
\end{equation}
Quite obviously, a similar analysis is possible in the general case
of a specific but arbitrary gauge group $G$.

\vspace{10pt}

\noindent{\bf Physical time evolution of the quantum system}. Let us 
now consider the physical time evolution of the system. According to
Ref.\cite{John1}, the corresponding operator is thus,
\begin{equation}
S_{\rm phys}(t_2,t_1)=e^{-i\hat{H}_0\,(t_2-t_1)}{\cal E}=
{\cal E}\,e^{-i\hat{H}_0(t_2-t_1)}\,{\cal E}\ \ \ ,
\label{eq:evol2}
\end{equation}
where the projection operator ${\cal E}$ onto the subspace
of physical states is defined by\footnote{In the present case,
the spectrum of $\hat{Q}^a$ being discrete, no $\delta$-limiting
procedure is required to properly define the reduced Hilbert space.},
\begin{equation}
{\cal E}=\int dU(\theta^a)\,e^{-i\theta^a\hat{Q}^a}\ \ \ .
\end{equation}
Here, $dU(\theta^a)$ is the Haar measure over the gauge group $G$,
the domain of integration being chosen according to the group $G$
rather than its universal covering group $G_{\rm univ}$ when different.
Once again, note that this measure is entirely specified by the requirement
of the properties in (\ref{eq:Proj1}) defining a projector,
thereby avoiding at once both issues of gauge fixing and of the
possibility of Gribov problems of the first or the second type\cite{Gov4}
related to a choice of gauge fixing.

Consider now the matrix element of the evolution operator between some
initial and final states, $|\psi_i>$ and $|\psi_f>$ respectively, for a time
interval $[t_i,t_f]$, namely,
\begin{equation}
P(i\rightarrow f)=<\psi_f|{\cal E}e^{-i\hat{H}_0(t_f-t_i)}
{\cal E}|\psi_i>\ \ \ .
\end{equation}
As seen previously, the states $|a_1,a_2,\cdots,a_n>$ 
in (\ref{eq:spacestates}) span
a complete orthonormalised basis of the space of states, including
gauge variant ones, whereas the subset 
$|n_1,\cdots,n_\ell>$ in (\ref{eq:physstates}) determines an orthonormalised
basis of the space of physical states. Therefore, one may write,
\begin{equation}
\begin{array}{c c l}
P(i\rightarrow f)&=&\sum_{n=0}^\infty\sum_{a_1,a_2,\cdots,a_n}
\sum_{m=0}^\infty\sum_{b_1,b_2,\cdots,b_m}
<\psi_f|a_1,a_2,\cdots,a_n>\times \\ \\
&\times &<a_1,a_2,\cdots,a_n|{\cal E}e^{-i\hat{H}_0(t_f-t_i)}
{\cal E}|b_1,b_2,\cdots,b_m>
<b_1,b_2,\cdots,b_m|\psi_i>\ \ \ .
\end{array}
\end{equation}
However, owing to the projection operators ${\cal E}$ to the left
and to the right of the exponentiated Hamiltonian operator, only
physical states do contribute to the sums over intermediate states.
In addition, these physical states diagonalise the Hamiltonian $\hat{H}_0$,
so that finally one obtains,
\begin{equation}
P(i\rightarrow f)=\sum_{n_1,\cdots,n_\ell=0}^\infty\,
e^{-im(n_1r_1+\cdots +n_\ell r_\ell+D_G/2)(t_f-t_i)}\,
<\psi_f|n_1,\cdots,n_\ell><n_1,\cdots,n_\ell|\psi_i>\ \ \ .
\label{eq:prop2}
\end{equation}

In conclusion, the physical evolution operator in (\ref{eq:evol2}) does indeed
propagate as intermediate states physical states only,
and in a manner which is consistent with the physical spectrum of the system.
Moreover, any unphysical component of the external states, which thus
has a vanishing overlap with the intermediate states $|n_1,\cdots,n_\ell>$, 
is not propagated by the physical evolution operator. In fact, the matrix
element $P(i\rightarrow f)$ vanishes identically whenever either one or
both of the external states does not possess a gauge invariant component.
It is not that gauge variant components of states are not propagated
in time in the system, but rather that the physical evolution operator
in (\ref{eq:evol2}) does not propagate the gauge variant component of states.

Given the general result in (\ref{eq:prop2}), note also that it is
possible in principle to compute any matrix element of the physical
evolution operator (\ref{eq:evol2}), given the appropriate choice
of initial and final states. For example, using the configuration space
wave functions of physical states such as those given 
in (\ref{eq:wavefunction}) in the case of $SU(2)$,
it is possible to obtain the configuration space matrix elements
of the physical evolution operator.
Another possible choice is that of phase space coherent states.

\vspace{10pt}

\noindent{\bf Phase space coherent states}. Finally, let us consider 
the phase space coherent states defined by,
\begin{equation}
|\pi^a,A^b;\eta>=e^{i\alpha(\pi^a,A^b)}\,e^{-iA^a\hat{\pi}^a}\,
e^{i\pi^a\hat{A}^a}\,|\eta>\ \ \ ,
\label{eq:coherent}
\end{equation}
where the choice of normalised fiducial state $|\eta>$ is arbitrary, as well
as the phase factor $\alpha(\pi^a,A^b)$.

Given the physical evolution operator in (\ref{eq:evol2}),
its phase space coherent state matrix elements are given by,
\begin{equation}
P(1\rightarrow 2)=<\pi_2,A_2;\eta|{\cal E}e^{-i\hat{H}_0(t_2-t_1)}{\cal E}
|\pi_1,A_1;\eta>\ \ \ .
\label{eq:coherent12}
\end{equation}
The evaluation of this expression requires the calculation of the
action of the projector ${\cal E}$ being applied to coherent states,
and more specifically the result for,
\begin{equation}
e^{-i\theta^a\hat{Q}^a}\,|\pi,A;\eta>\ \ \ .
\end{equation}

Given the parameters $\theta^a$ and the degrees of freedom $\pi^a$
and $A^a$, let us define the quantities $\pi^a_\theta$ and
$A^a_\theta$ by the relations,
\begin{equation}
A^a_\theta T^a=U A^aT^a U^{-1}\ \ \ ,\ \ \ 
\pi^a_\theta T^a=U \pi^a T^a U^{-1}\ \ \ ,
\end{equation}
where $U$ is the finite gauge transformation in the group $G$,
\begin{equation}
U=e^{-i\theta^a T^a}\ \ \ .
\end{equation}
Then, it is possible to show that one has,
\begin{equation}
e^{-i\theta^a\hat{Q}^a}\,|\pi^a,A^b;\eta>=
|\pi^a_\theta,A^b_\theta;\eta_\theta>\ \ \ ,
\end{equation}
where the coherent state on the r.h.s. is defined as in (\ref{eq:coherent})
with the fiducial state $|\eta_\theta>$ now given by,
\begin{equation}
|\eta_\theta>=e^{-i\theta^a\hat{Q}^a}\,|\eta>\ \ \ .
\end{equation}

Consequently, the phase space coherent space matrix element
in (\ref{eq:coherent12}) takes the form,
\begin{equation}
\begin{array}{r l}
&P(1\rightarrow 2)=\\ \\
&=\int dU(\theta^a_2) \int dU(\theta^a_1) \,
<(\pi_2)^a_{\theta_2},(A_2)^a_{\theta_2};\eta_{\theta_2}|
e^{-\frac{1}{2}i(t_2-t_1)[(\hat{\pi}^a)^2+m^2(\hat{A}^a)^2]}
|(\pi_1)^a_{\theta_1},(A_1)^a_{\theta_1};\eta_{\theta_1}>\ \ .
\end{array}
\label{eq:prop4}
\end{equation}
In view of the integration over the group parameters $\theta^a_1$
and $\theta^a_2$, it would be reasonable to believe that only
gauge invariant physical states contribute to this expression as
intermediate states. As shown in (\ref{eq:prop2}), this is indeed
the case, and the correct spectrum of physical states is in fact
recovered from the time dependence of this expression.

One may attempt, which shall not be done here, to compute (\ref{eq:prop4})
explicitly. This should be particularly simple for the choice
$|\eta>=|0>$, in which case $|\eta_\theta>=|0>$ as well. However,
the expression in (\ref{eq:prop2}) seems to be better
suited for the purpose of a calculation of (\ref{eq:prop4}), 
since one then only requires the overlap
functions
\begin{equation}
<n_1,\cdots,n_\ell|\pi^a,A^b;\eta>\ \ \ ,
\end{equation}
of the phase space coherent states with the physical states 
$|n_1,\cdots,n_\ell>$.
These functions may be obtained using the Fock representation
of the operators $\hat{A}^a$ and $\hat{\pi}^a$.

\section{Conclusions}
\label{Sect5}

Klauder's proposal\cite{John1} for the reproducing kernel or propagator of
physical states in gauge invariant systems is based on the
physical projector ${\cal E}$ onto the reduced Hilbert space 
of physical states. As shown in Ref.\cite{John1},
the path integral measure for the Lagrange multipliers associated
to the constraints is then uniquely determined from the projector
property ${\cal E}^2={\cal E}$ of this operator, independently
of any gauge fixing conditions or reduction of second class constraints.
In addition, Klauder's approach does not require the introduction 
in the path integral representation of gauge invariant observables
of the $\delta$-functionals and functional determinants
which are characteristic of the conventional
approaches to the quantisation of constrained systems.

In the present letter, it is pointed out that
since Klauder's physical propagator does not necessite gauge
fixing conditions, potential Gribov problems, which are characteristic
of the conventional approaches to constrained systems, are avoided
from the outset, while the properties of the physical projector ${\cal E}$
also ensure that Klauder's physical propagator does indeed
lead to the correct physically consistent results for gauge invariant
observables, by effectively including once and only once
the contribution from each of the inequivalent gauge orbits of the system,
as would result from an admissible choice of gauge fixing conditions
in the conventional approaches.
In other words, the role of the physical projector ${\cal E}$ is also
to effectively determine the physically consistent integration measure
over the modular space of the system---{\sl i.e.\/} the quotient of 
configuration space or phase space, including Lagrange multiplier variables,
by the gauge group.
This important aspect of Klauder's proposal is confirmed explicitly
in all these aspects by two simple examples, namely the free relativistic 
scalar particle and pure Yang-Mills theory in 0+1 dimensions.

The analysis is performed within the abstract operator formulation of
a quantised constrained system with first class constraints only whose
algebra is closed. Klauder's original discussion\cite{John1} is presented
within the context of the phase space coherent state path integral
quantisation of constrained systems. As is well know, the operator
approach can be used to develop and justify the path integral one,
thereby specifying unambiguously the integration measures over 
the phase space degrees of freedom and Lagrange multiplier variables
in as far as the quantised system itself is uniquely and well defined
at the operator level. In addition, the formulation of the physical
projection operator ${\cal E}$ is such that manifest gauge invariance
and covariance under other specific symmetries that the system may possess
is maintained throughout. There is no need to develop a BRST description
with its additional auxiliary and ghost degrees of freedom to achieve that aim,
while the BRST approach is usually also fraught with Gribov problems.

Clearly, it would be extremely interesting to apply Klauder's point of view
to other gauge invariant systems of physical interest, and see how
the corresponding results compare to the understanding which has
developed on the basis of the conventional approaches.
Before considering more realistic theories in 3+1 dimensions,
obvious candidates would be Yang-Mills and Chern-Simons theories,
as well as quantum gravity theories, in
1+1 and 2+1 dimensions, the quantum gravity theories in 1+1 dimensions
including of course string theories.

\section*{Acknowledgements}

Prof. J. Klauder is gratefully acknowledged for very useful discussions
and remarks concerning the present work and Ref.\cite{John1}.
Dr.~S.V.~Shabanov is thanked for calling Ref.\cite{Shabanov1}
to the author's attention, as well as pointing out that the projection
operator approach has already been considered earlier\cite{Shabanov2}.


\clearpage

\begin{thebibliography}{99}

\bibitem{John1} J.R.~Klauder, {\sl Coherent State Quantization of 
Constraint Systems\/},\\
preprint {\tt quant-ph/9604033} (April 1996).

\bibitem{Coherent} For a review, see for example,\\
J.R.~Klauder and B.-S.~Skagerstam, {\sl Coherent States:
Applications in Physics and Mathematical Physics\/,} (World Scientific,
Singapore, 1985).

\bibitem{Gribov} V.N.~Gribov, {\sl Nucl. Phys.} {\bf B139} (1978) 1.

\bibitem{Singer} I.M.~Singer, {\sl Comm. Math.} {\bf 60} (1978) 7.

\bibitem{Gov4} For a general review and explicit examples, see\\
J.~Govaerts, {\sl Hamiltonian Quantisation and Constrained
Dynamics} (Leuven University Press, Leuven, 1991).

\bibitem{Hen1} For a review and references to the original
literature, see also,\\
M.~Henneaux, {\sl Physics Reports} {\bf 126} (1985) 1;\\
M.~Henneaux and C.~Teitelboim, {\sl Quantization of Gauge
Systems\/} (Princeton University Press, Princeton, New Jersey, 1992).

\bibitem{Faddeev} L.D.~Faddeev, {\sl Theor. Math. Phys.} {\bf 1} (1970) 1.

\bibitem{Dirac} P.A.M.~Dirac, {\sl Lectures on Quantum Mechanics}
(Belfer Graduate School of Science, Yeshiva University, New York, 1964).

\bibitem{FV} E.S.~Fradkin and G.A.~Vilkovisky, {\sl Phys. Lett.}
{\bf B55} (1975) 224;\\
I.A.~Batalin and G.A.~Vilkovisky, {\sl Phys. Lett.} {\bf B69} (1977) 309;\\
E.S.~Fradkin and T.E.~Fradkina, {\sl Phys. Lett.} {\bf B72} (1978) 343;\\
I.A.~Batalin and E.S. Fradkin, {\sl Rivista del Nuovo Cimento}
{\bf 9} (1986) 1.

\bibitem{Gov1} J.~Govaerts, {\sl Int. J. Mod. Phys.} {\bf A4} (1989) 173.

\bibitem{Gov2} J.~Govaerts, {\sl Int. J. Mod. Phys.} {\bf A4} (1989) 4487.

\bibitem{Gov3} J. Govaerts and W.~Troost, {\sl Class. Quantum Grav.}
{\bf 8} (1991) 1723.

\bibitem{Feynman} R.P.~Feynman, {\sl Acta Phys. Pol.} {\bf 24} (1963) 697.

\bibitem{Hajicek} P.~Hacijek, {\sl J. Math. Phys.} {\bf 27} (1986) 1800.

\bibitem{Gov5} For a detailed analysis of this system, see Ref.\cite{Gov4}.
Some original material is to be found 
in Refs.\cite{Teitelboim,HenTeitel,Monaghan}.

\bibitem{Teitelboim} C.~Teitelboim, {\sl Phys. Rev.} {\bf D25} (1982) 3159.

\bibitem{HenTeitel} M.~Henneaux and C.~Teitelboim, {\sl Ann. Phys.} {\bf 143}
(1982) 127.

\bibitem{Monaghan} S.~Monaghan, {\sl Phys. Lett.} {\bf B178} (1986) 231.

\bibitem{Shabanov1} L.V.~Prokhorov and S.V.~Shabanov, {\sl Phys. Lett.} 
{\bf B216} (1989) 341.

\bibitem{Shabanov2} L.V.~Prokhorov and S.V.~Shabanov, 
{\sl Sov. Phys. Uspekhi\/} {\bf 34} (1991) 108, and references therein.

\end{thebibliography}
\end{document}